\documentclass[aps,prl,superscriptaddress,showpacs,amssymb,amsmath,amsfonts,twocolumn]{revtex4}
\usepackage{graphicx}

\begin{document}

\title{$\eta{'}$ photoproduction on the proton for photon energies
from 1.527 to 2.227 GeV}

\newcommand*{\ASU}{Arizona State University, Tempe, Arizona 85287-1504}
\affiliation{\ASU}
\newcommand*{\UCLA}{University of California at Los Angeles, Los Angeles, California  90095-1547}
\affiliation{\UCLA}
\newcommand*{\CSUDH}{California State University, Dominguez Hills, Carson, California 90747-0005}
\affiliation{\CSUDH}
\newcommand*{\CMU}{Carnegie Mellon University, Pittsburgh, Pennsylvania 15213}
\affiliation{\CMU}
\newcommand*{\CUA}{Catholic University of America, Washington, D.C. 20064}
\affiliation{\CUA}
\newcommand*{\SACLAY}{CEA-Saclay, Service de Physique Nucl\'eaire, F91191 Gif-sur-Yvette,Cedex, France}
\affiliation{\SACLAY}
\newcommand*{\CNU}{Christopher Newport University, Newport News, Virginia 23606}
\affiliation{\CNU}
\newcommand*{\UCONN}{University of Connecticut, Storrs, Connecticut 06269}
\affiliation{\UCONN}
\newcommand*{\DUKE}{Duke University, Durham, North Carolina 27708-0305}
\affiliation{\DUKE}
\newcommand*{\ECOSSEE}{Edinburgh University, Edinburgh EH9 3JZ, United Kingdom}
\affiliation{\ECOSSEE}
\newcommand*{\FIU}{Florida International University, Miami, Florida 33199}
\affiliation{\FIU}
\newcommand*{\FSU}{Florida State University, Tallahassee, Florida 32306}
\affiliation{\FSU}
\newcommand*{\GWU}{The George Washington University, Washington, DC 20052}
\affiliation{\GWU}
\newcommand*{\ECOSSEG}{University of Glasgow, Glasgow G12 8QQ, United Kingdom}
\affiliation{\ECOSSEG}
\newcommand*{\GEISSEN}{Physikalisches Institut der Universitaet Giessen, 35392 Giessen, Germany}
\affiliation{\GEISSEN}
\newcommand*{\ISU}{Idaho State University, Pocatello, Idaho 83209}
\affiliation{\ISU}
\newcommand*{\INFNFR}{INFN, Laboratori Nazionali di Frascati, 00044, Frascati, Italy}
\affiliation{\INFNFR}
\newcommand*{\INFNGE}{INFN, Sezione di Genova, 16146 Genova, Italy}
\affiliation{\INFNGE}
\newcommand*{\ORSAY}{Institut de Physique Nucleaire ORSAY, Orsay, France}
\affiliation{\ORSAY}
\newcommand*{\ITEP}{Institute of Theoretical and Experimental Physics, Moscow, 117259, Russia}
\affiliation{\ITEP}
\newcommand*{\JMU}{James Madison University, Harrisonburg, Virginia 22807}
\affiliation{\JMU}
\newcommand*{\KYUNGPOOK}{Kyungpook National University, Daegu 702-701, South Korea}
\affiliation{\KYUNGPOOK}
\newcommand*{\MIT}{Massachusetts Institute of Technology, Cambridge, Massachusetts  02139-4307}
\affiliation{\MIT}
\newcommand*{\UMASS}{University of Massachusetts, Amherst, Massachusetts  01003}
\affiliation{\UMASS}
\newcommand*{\MOSCOW}{Moscow State University, General Nuclear Physics Institute, 119899 Moscow, Russia}
\affiliation{\MOSCOW}
\newcommand*{\UNH}{University of New Hampshire, Durham, New Hampshire 03824-3568}
\affiliation{\UNH}
\newcommand*{\NSU}{Norfolk State University, Norfolk, Virginia 23504}
\affiliation{\NSU}
\newcommand*{\OHIOU}{Ohio University, Athens, Ohio  45701}
\affiliation{\OHIOU}
\newcommand*{\ODU}{Old Dominion University, Norfolk, Virginia 23529}
\affiliation{\ODU}
\newcommand*{\PITT}{University of Pittsburgh, Pittsburgh, Pennsylvania 15260}
\affiliation{\PITT}
\newcommand*{\RPI}{Rensselaer Polytechnic Institute, Troy, New York 12180-3590}
\affiliation{\RPI}
\newcommand*{\RICE}{Rice University, Houston, Texas 77005-1892}
\affiliation{\RICE}
\newcommand*{\URICH}{University of Richmond, Richmond, Virginia 23173}
\affiliation{\URICH}
\newcommand*{\SCAROLINA}{University of South Carolina, Columbia, South Carolina 29208}
\affiliation{\SCAROLINA}
\newcommand*{\JLAB}{Thomas Jefferson National Accelerator Facility, Newport News, Virginia 23606}
\affiliation{\JLAB}
\newcommand*{\UNIONC}{Union College, Schenectady, NY 12308}
\affiliation{\UNIONC}
\newcommand*{\VT}{Virginia Polytechnic Institute and State University, Blacksburg, Virginia   24061-0435}
\affiliation{\VT}
\newcommand*{\VIRGINIA}{University of Virginia, Charlottesville, Virginia 22901}
\affiliation{\VIRGINIA}
\newcommand*{\WM}{College of William and Mary, Williamsburg, Virginia 23187-8795}
\affiliation{\WM}
\newcommand*{\YEREVAN}{Yerevan Physics Institute, 375036 Yerevan, Armenia}
\affiliation{\YEREVAN}

\author {M. Dugger}
\affiliation{\ASU}
\author {J.P.~Ball}
\affiliation{\ASU}
\author {P. ~Collins}
\affiliation{\ASU}
\author {E.~Pasyuk}
\affiliation{\ASU}
\author {B.G.~Ritchie}
\affiliation{\ASU}
\author {G.~Adams} 
\affiliation{\RPI}
\author {P.~Ambrozewicz} 
\affiliation{\FIU}
\author {E.~Anciant} 
\affiliation{\SACLAY}
\author {M.~Anghinolfi} 
\affiliation{\INFNGE}
\author {B.~Asavapibhop} 
\affiliation{\UMASS}
% ...
\collaboration{The CLAS Collaboration}
     \noaffiliation

\date{\today}

\begin{abstract}

Differential cross sections for the reaction $\gamma p \rightarrow
\eta{'} p$ have been measured with the CLAS spectrometer and a
tagged photon beam with energies from 1.527 to 2.227 GeV. 
The results reported here possess much greater accuracy than previous measurements. 
Analyses of these data suggest for the first time the coupling of the
$\eta{'}N$ channel to both the $S_{11}(1535)$ and $P_{11}(1710)$ resonances, known to couple 
strongly to the $\eta{N}$ channel in photoproduction on the proton,
and the importance of $J=3/2$ resonances in the process.

\end{abstract}


\maketitle

Understanding the structure of the proton is challenging due to
the great complexity of this strongly interacting multi-quark system \cite{Cap00}. 
Of particular utility in investigating nucleon structure are those production
mechanisms and observables that help isolate 
individual excited states of the nucleon and 
determine the importance of specific contributions. Since the
electromagnetic interaction is well understood, photoproduction offers
one of the more powerful methods for studying the nucleon. 
Since the $\eta$ and $\eta{'}$ mesons have isospin 0, $\eta N$ and $\eta{'} N$
final states can only originate (in one-step processes) from isospin $I = 1/2$ 
intermediate states. Therefore, the reactions $\gamma p \rightarrow \eta p$ and
$\gamma p -> \eta{'} p$ isolate $I = 1/2$ resonances, thereby providing an
``isospin filter'' for the spectrum of broad, overlapping nucleon
resonances, a useful simplification for theoretical efforts to predict
the large number of excited nucleon states.

Thus, photoproduction of the $\eta{'}$ meson from the proton is an
excellent tool for clarifying the details of the nucleon resonance spectrum. 
However, existing data for the $\gamma p \to
\eta{'} p$ reaction come from only a few 
exclusive or semi-exclusive measurements
due to the limitations of experimental facilities. 
While previous experiments \cite{abbhhm,ahhm,saphir} detected fewer than 300 $\eta{'}$ events,
in the measurements described here, 
over 2$\times 10^5$ $\eta{'}$ photoproduction events 
were detected and used to extract differential cross sections.  

The differential cross sections for the reaction $\gamma p \rightarrow
\eta{'} p$ were measured with 
the CEBAF Large Acceptance Spectrometer (CLAS) \cite{CLAS} 
and the bremsstrahlung photon tagging facility \cite{tagger}  
at the Thomas Jefferson National Accelerator Facility.
The cross sections were part of a program of meson production measurements
using the same CLAS, tagger, and target configuration.  
Tagged photons, with energies $E_\gamma$ between 0.49 and 2.96 GeV, 
were incident on an
18-cm-long liquid hydrogen target placed at the center of CLAS. 
(The threshold for $\eta{'}$ photoproduction on the proton is $E_\gamma$ = 1.447 GeV.)
The event trigger required the coincidence of a post-bremsstrahlung
electron passing through the focal plane of the photon tagger and at
least one charged particle detected in CLAS.  Tracking of the charged
particles through the magnetic field within CLAS by drift chambers
provided determination of their charge, momentum and scattering
angle. This information, together with the particle velocity measured
by the time-of-flight scintillators, provided particle identification
for each particle detected in CLAS and its corresponding momentum four-vector.
Particle identification was generally unambiguous; in the case
of proton identification, the fraction of particles misidentified as protons made up
a background of less than $2\times10^{-3}$. 

\begin{figure}

\includegraphics * [scale=0.35]{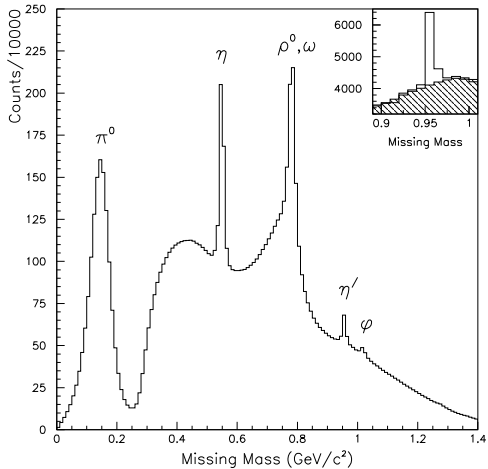}
\caption{Missing mass spectrum for $\gamma p \rightarrow p X$ 
integrated over all photon energies and angles.  
Inset: Missing mass spectrum binned in photon energy
(1.728 $\pm$ 0.025 GeV) and angle (-0.2 $\leq \cos\vartheta^{\eta{'}}_{\rm{c.m.}}$
$\leq$ 0.0), illustrating a typical bin. 
The shaded area shows the 
multi-pion background discussed in the
text.\label{mm}}
\end{figure}

The $\gamma p \to p X$ missing mass was used to identify 
photoproduced mesons through detection of the proton recoiling into the CLAS
from the cryogenic target. 
As seen in the missing mass spectrum
in Fig. 1, the resolution obtained is sufficient for clear
identification of the photoproduced $\pi^0, \eta, \rho + \omega, \eta{'}$, and 
$\varphi$ mesons, the latter four peaks situated 
atop a multi-pion background.
The missing mass spectrum was binned in center-of-mass
scattering angle and photon energy to extract meson yields for each angle/energy bin.  
The CLAS acceptance limited
the measurement of the $\gamma p \to \eta{'}p $ reaction to photon energies
above 1.527 GeV ($W$ = 1.94 GeV) 
and $\eta{'}$ center-of-mass scattering angles $\vartheta^{\eta{'}}_{\rm{c.m.}}$
in the range $-0.8 \leq \cos\vartheta^{\eta{'}}_{\rm{c.m.}} \leq 0.8$.
For the $\eta{'}$ measurements reported here,
a total of 15 non-overlapping bins in incident photon energy $E_\gamma$ were used, each about 50 MeV wide. 
(For convenience, the photon energy bins are labeled by the energy of the centroid of the bin.)
The photon energies ranged from bins centered on $E_\gamma$ from 1.527 up to 2.227 GeV,
corresponding to center-of-mass energies $W$ from 1.94 to 2.25 GeV.
Above this energy range, the yield for $\eta{'}$ photoproduction was
too low to permit the extraction of reliable cross sections. 
The background subtraction (as exemplified in the inset
in Fig.\ \ref{mm}) assumed a mixture of two-,
three-, and four-pion contributions, along with contributions from the
$\rho^0$.  

The proton detection efficiency for CLAS was measured empirically 
\cite{Dug01,Aug99} 
using the reaction $\gamma p \rightarrow p \pi^+ \pi^-$.  
Both pions were required to be detected in the
event and both must have been produced by the same
photon from the bremsstrahlung beam.  A missing mass reconstruction
from the kinematical information for the two pions was performed to
determine if a proton should have been seen in the CLAS in a
particular phase-space volume.  The presence
or absence of a proton yielded an empirical measure of
the momentum-dependent proton detection efficiency for that volume.
Efficiency uncertainties for $\eta{'}$ photoproduction, dominated by
the statistical uncertainty in the number of protons scattered and
detected, were determined for each bin, and ranged from
$\sim$1\% at the lowest energies to $\sim$2\% at the highest energies.

The results reported here represent the first measurements for $\eta{'}$
photoproduction utilizing an {\em absolute} measurement of the photon flux \cite{gflux}.  
The photon flux for the entire tagger photon energy range was determined by measuring
the rate of scattered electrons detected in each counter of the focal plane of
the bremsstrahlung photon tagger by sampling focal plane hits 
not in coincidence with CLAS.
The detection rate for the scattered electrons
was integrated over the livetime of the experiment
and converted to the total number of photons on target
for each counter of the tagger focal plane.
The tagging efficiency was
measured in dedicated runs with a Total Absorption Counter (TAC) \cite{tagger},
which directly counted all photons in the beam \cite{gflux}.

Ideally, one would use a well-known reaction 
in the energy range used for these measurements
to confirm the validity of the photon flux measurement technique and 
to estimate the uncertainties in the photon flux normalization.
However, no large database exists for any 
photoproduction reaction over the range of photon energies
for which we report $\eta{'}$ cross sections here.
As an alternative, the pion photoproduction database is quite extensive.
The SAID parameterization \cite{SAID} provides a very good description of that
database. The SAID analysis incorporates many observables for all
channels of pion photoproduction. The existing $\pi^0$ photoproduction
cross section database below 1.5 GeV is quite dense. 
(The data below 1.5 GeV make up 95\% of the published measurements on $\pi^0$ 
photoproduction on the proton.)
The SAID solution (SM02) is in 
very good agreement with those existing data. Thus, SAID can be assumed
to provide the correct energy and angular dependence for the $\pi^0$
photoproduction cross section in that energy range within its estimated
normalization uncertainty of 2\%. The existing data above 1.5 GeV are
much more scarce and have significantly larger uncertainties. 
Therefore, we have used that parametrization 
to ascertain the validity of the procedures used here 
by comparing that SAID parametrization 
to $\pi^0$ photoproduction cross sections (for $E_\gamma$ from 0.675 to 1.525 GeV) 
extracted from data taken
simultaneously with the $\eta{'}$ measurements reported here 
(for $E_\gamma$ from 1.527 to 2.227 GeV), 
using the same absolute normalization techniques for both reactions \cite{pi0_soon}.   

In order to determine the $\pi^0$ cross sections for this experiment 
over the photon energy range from 0.675 to 1.525 GeV, 
the empirically measured proton detection efficiency for
CLAS had to be supplemented by a Monte-Carlo estimate of the detection
efficiency for protons from $\pi^0$ photoproduction because the phase
space occupation of protons for the $\gamma p \rightarrow p \pi^+
\pi^-$ reaction becomes sparse at higher energies when rebinned for
$\gamma p \rightarrow p \pi^0$ efficiencies.  Agreement between
the empirical and Monte-Carlo methods, where sufficient statistics made 
comparison possible, was within 3\%.

For $E_\gamma$ from 0.675 to 1.525 GeV 
and the range of $\cos(\vartheta^{\pi^0}_{\rm{c.m.}})$ used here,
our entire set of $\pi^0$ differential cross sections, 
comprised of 19
energy bins each with 12 bins in $\cos(\vartheta^{\pi^0}_{\rm{c.m.}})$ (228
points, in total) was easily fit by the SAID parametrization with a single
overall constant factor $N$ = 1.02 ($\chi^2_{reduced} = 1.3$).  {\em{This overall
agreement throughout the energy range
implies that the absolute normalization technique is sound, 
and additionally indicates the detector acceptance also is well-determined.}}

To estimate the uncertainty in
the photon flux measurement, a more refined fit of our measured 
differential cross sections for
$\pi^0$ photoproduction for {\em{each}} photon energy bin 
to the SAID parametrization was
performed, determining a single overall constant factor $N_E$ for {\em{each}}
photon energy bin. For $E_\gamma$ = 0.675 to 1.525 GeV,
these $N_E(E_\gamma)$ values were produced, binned into a histogram, and fit
with a simple Gaussian.  The centroid of the fit to $N_E(E_\gamma)$ was
1.02, as before. The standard
deviation $\sigma(N_E(E_\gamma))$ of the $N_E(E_\gamma)$ values was 4\%.  
We conservatively estimate the absolute normalization systematic uncertainty 
to be about 5\%.

\begin{figure}
\includegraphics[scale=0.35]{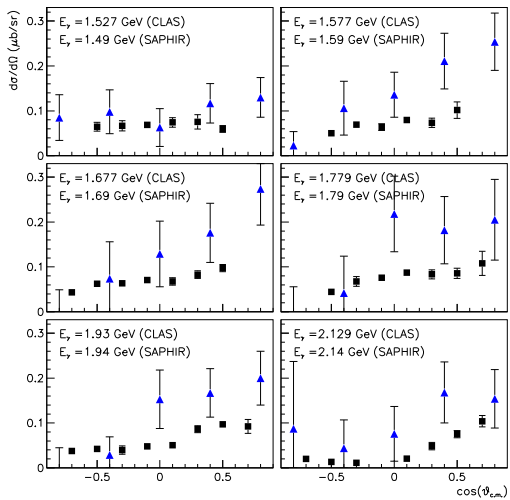}
\caption{Differential cross sections for $\eta{'}$ photoproduction on
the proton (black squares).  Other results from SAPHIR \cite{saphir}
(blue triangles) are shown for comparison. Error bars shown include systematic and
statistical uncertainties.
\label{diff_pic}}
\end{figure}

\begin{figure}
\includegraphics[scale=0.35]{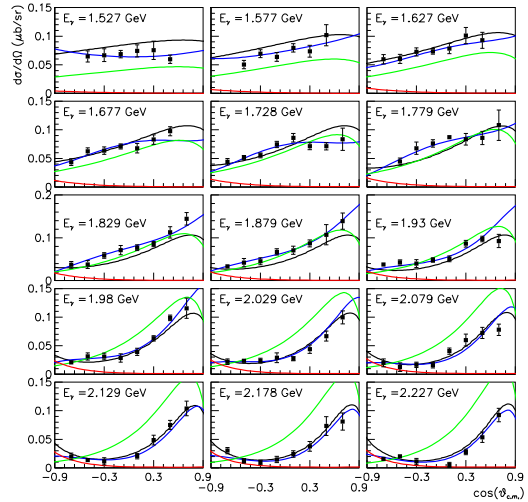}
\caption{Differential cross sections for $\eta{'}$ photoproduction on
the proton. Also shown are results from Nakayama and Haberzettl
\cite{hk} (Red lines: $u$-channel contributions. 
Green lines: $t$-channel contributions. Blue lines: Sum of all $s$-, $t$-, and $u$-channel 
contributions), and a model (black lines) inspired by
A. Sibirtsev \textit{et al.} \cite{elster_and_co}, as discussed in the
text. Error bars shown include systematic and
statistical uncertainties.
\label{diff_comp}}
\end{figure}

The differential cross sections for $\eta{'}$
photoproduction obtained are shown in Figs.\ \ref{diff_pic} and
\ref{diff_comp}.
In general, the angular distributions, while flat at
threshold, show a continuing increase in slope at forward angles with
increasing photon energy. At the highest energies, growth at backward
angles is also seen. These general features are suggestive of coupling
to an s-channel resonance near threshold, with increasing contributions
of t- and u-channel exchange as the energy above threshold increases.
The SAPHIR measurements \cite{saphir} are shown
for comparison in Fig.\ \ref{diff_pic}.  The CLAS data, with much
smaller error bars and smaller photon energy bins
(SAPHIR has energy bins of 100 MeV for energies below 1.84 GeV and 200 MeV wide bins above),
generally agree with the SAPHIR results
within the very large error bars of the latter, but the CLAS values
are nonetheless systematically lower.  The excellent agreement
noted above between
the SAID parametrization and the $\pi^0$ photoproduction cross sections
measured here, using the same normalization techniques as used for these $\eta{'}$
cross sections, strongly suggests the absolute normalization determined here is
correct.

Included in Fig.\ \ref{diff_comp} are the results (shown as red, green, and blue
lines) representing a consistent analysis of the reactions $\gamma p
\rightarrow p \eta{'}$ and $p p \rightarrow p p \eta{'}$ by Nakayama
and Haberzettl (NH) \cite{hk}.  
The NH analysis is based upon a relativistic
meson-exchange model of hadronic interactions including coupled-production mechanisms.
We have also performed 
calculations (black lines) using a relativistic
meson-exchange model by A. Sibirtsev
\textit{et al.} \cite{elster_and_co} as a recipe.  For both models,
allowed processes include $s$-, $t$-, and $u$-channel contributions.
The intermediate mesons in the $t-$channel exchanges 
are the $\omega$ and $\rho^0$ in both cases. Both models here also included the
$S_{11}(1535)$ and $P_{11}(1710)$ resonances ($J = 1/2$), which are 
known to decay strongly to the ${\eta}N$ channel
\cite{PDG}.  The NH model also includes two additional $S_{11}$ and two additional 
$P_{11}$ resonances, albeit with relatively small couplings. In contrast to the fit 
of the SAPHIR data in Ref. \cite{saphir}, the present adaptation of the NH model to 
our data now also requires $J = 3/2$ resonances [$P_{13}(1940)$, $D_{13}(1780)$, and $D_{13}(2090)$].
Since the NH model fits the data better than our calculations, 
the inclusion of these additional $J = 3/2$ resonances appears to be beneficial. 

A comparison of the predictions of these two different approaches can provide
insight into which physical contributions are most successful
at explaining features of the observed cross sections. 
The forward peaking of the cross sections at the highest energies
is dominated by $t$-channel exchange.  
Addition of the $S_{11}(1535)$ state contributes mainly to the
overall initial rise and fall of the total cross sections below 1.7 GeV.
We note that this is 
the first time that $S_{11}(1535)$ and $P_{11}(1710)$ resonances, known to strongly
couple to the $\eta{N}$ channel in photoproduction, have been used in fits as contributions to
the $\eta{'}N$ photoproduction channel. The $J=3/2$ resonances included by NH are
especially useful in obtaining the correct shape of the differential
cross sections for the energies from
1.728 to 1.879 GeV.  The $u$-channel exchange causes the backward-angle 
enhancement seen around 2 GeV and above. 
(The general behavior of individual $t-$ and $u-$channel
contributions can be seen in Fig. \ref{diff_comp} and Ref. \cite{hk2}.)

Since the $\eta{'}$ meson is the only flavor singlet of the
fundamental pseudoscalar meson nonet, studies of the reaction can also help yield
information on the role of glue states in excitations of the nucleon.
The flavor-singlet axial charge of the nucleon ($G_A(0)$) is related to
the $\eta{'}$-nucleon-nucleon and gluon-nucleon-nucleon coupling
constants ($g_{\eta{'}NN}$ and $g_{GNN}(0)$, respectively) through the
flavor-singlet Goldberger-Treiman relation \cite{ven}:
\begin{equation}
2 m_N G_A(0) = F g_{\eta{'}NN} - \frac{F^2 m_{\eta{'}}^2}{N_F} g_{GNN}(0), \label{eq:ga0}
\end{equation}
where $m_N$ is the mass of the nucleon, $m_{\eta{'}}$ is the $\eta{'}$ mass, 
$F$ is an invariant decay constant that reduces to $F_{\pi}$ (pion decay constant) 
if the $U(1)_A$ anomaly were turned off \cite{feld}, and $N_F$ equals the number of flavors.
When first measured \cite{emc}, the singlet axial charge was found to
have a value of $G_A(0) = 0.20 \pm 0.35$. (A more recent calculation \cite{hirai}
gives $G_A(0) = 0.213 \pm 0.138$.)
At that time, the importance of the second term in Eq.\ \ref{eq:ga0} was unappreciated, 
and this low value of $G_A(0)$ was surprising: Since $g_{\eta{'}NN}$
is considered to be correlated with the 
fraction of the nucleon spin carried by its constituent quarks
\cite{elster_and_co}, that fraction would then be consistent with zero. 
Thus, neglecting the gluonic portion of Eq.\ \ref{eq:ga0}
was one of the causes of the so-called ``spin
crisis."
However, by including the gluonic degrees of freedom in Eq.\ \ref{eq:ga0}, the
value of $g_{\eta{'}NN}$ can be large, provided that it is nearly canceled
by $g_{GNN}(0)$.  This equation then can be used to indirectly determine
the gluonic coupling to the nucleon given a value of $g_{\eta{'}NN}$.

The observed $u$-channel contribution seen here allows the $g_{\eta{'}NN}$
coupling to be extracted (albeit in a model-dependent way). The
value of $g_{\eta{'}NN}$ found from the particular NH fit shown here is 1.33, whereas
our results using the model of Ref.\ \cite{elster_and_co} provides 1.46. Since differential cross sections alone do not provide sufficient
constraints to these models, these $g_{\eta{'}NN}$ values
should be taken with caution. 
Nonetheless, both values of $g_{\eta{'}NN}$ are consistent with the analysis of 
Ref. \cite{feld} which gives 1.4 $\pm$ 1.1. Moreover, even though
the uncertainty in $g_{\eta{'}NN}$ precludes a definitive statement 
about the value of $g_{GNN}(0)$, Eq.\ \ref{eq:ga0} can be carried out taking $N_F = 3$, 
$F = F_{\pi} = 0.131$ GeV, $g_{\eta{'}NN} = 1.4$, and $G_A(0) = 2.13$, 
yielding the result that $g_{GNN}(0) = 41$ GeV$^{-3}$.

In conclusion, the differential cross sections presented here 
are the first high-quality data for the
$\gamma p \rightarrow \eta{'} p$ reaction.  
An analysis of the data with two different models of the process suggests for the first time 
contributions from both the $S_{11}(1535)$ and $P_{11}(1710)$ nucleon 
resonances to the $\eta{'}N$ channel in photoproduction, the two resonances
previously identified as strongly coupling to the $\eta{N}$ channel \cite{PDG}.
Using two different theoretical descriptions of the data, 
these cross sections suggest a value for the $\eta{'}$-nucleon-nucleon coupling
constant $g_{\eta{'}NN}$ of 1.3-1.5, 
consistent with previous theoretical estimates of this quantity.
These data should continue to prove quite useful in guiding future experimental  
and theoretical investigations of the structure of the nucleon.

The authors gratefully acknowledge the Jefferson Lab Accelerator
Division staff.  We thank W. Kaufmann, H. Haberzettl and K. Nakayama
for useful discussions and assistance.  This work was supported by the
National Science Foundation, the Department of Energy (DOE), the
Deutsche Forschungsgemeinschaft (through an Emmy Noether grant to
U.T.), the French Centre National de la Recherche Scientifique
and Commissariat \`a l'Energie Atomique, 
the Italian Istituto Nazionale di Fisica Nucleare,
and the Korean Science and Engineering Foundation.  
The Southeastern Universities Research
Association operates Jefferson Lab for DOE under contract
DE-AC05-84ER40150.


\begin{thebibliography}{}

\bibitem{Cap00} See, e.g., S. Capstick and W. Roberts, Prog. in Part. 
and Nucl. Phys {\textbf{45}}, S241 (2000).

\bibitem{abbhhm} ABBHHM Collaboration, Phys. Rev. {\textbf{175}}, 1669
(1968).

\bibitem{ahhm} AHHM Collaboration, Nucl. Phys. B {\textbf{108}}, 45
(1976).

\bibitem{saphir} R. Pl\"otzke {\textit{et al.}}, Phys. Lett. B
{\textbf{444}}, 555 (1998).

\bibitem{CLAS} B.~Mecking {\textit {et al.}},
Nucl. Instr. Meth. {\textbf A 503}, 513 (2003).

\bibitem{tagger} D. Sober {\textit {et al.}},
Nucl. Instr. Meth. {\textbf A 440}, 263 (2000).

\bibitem{Dug01} M. Dugger, Ph.D. dissertation, Arizona State
University (unpublished, 2001).

\bibitem{Aug99} T. Auger, Ph.D. dissertation, Universit\'e de Paris
(unpublished, 1999).

\bibitem{gflux} J. Ball and E. Pasyuk, CLAS note 2005-002 (2005)
http://www.jlab.org/Hall-B/notes/clas\_notes05/2005-002.pdf

\bibitem{SAID} R. A. Arndt, W. J. Briscoe, I. I. Strakovsky, and
R. L. Workman, Phys. Rev. C {\textbf{66}}, 055213 (2002); 
R. A. Arndt, I. I. Strakovsky, R. L. Workman, and M. M. Pavan, Phys. Rev. C
\textbf{52}, 2120 (1995).

\bibitem{pi0_soon} M. Dugger, {\textit{et al.}}, The results for $\pi^0$ 
photoproduction will be reported in a forthcoming publication.

\bibitem{hk} K. Nakayama, and H. Haberzettl, private communication. The 
results shown here are based on an extended version of the model given in 
\protect{\cite{hk2}}. 

\bibitem{PDG} S. Eidelman {\textit{et al.}}, Phys. Lett. B {\textbf{592}}, 1 (2004).

\bibitem{hk2} K. Nakayama and H. Haberzettl, Phys. Rev. C {\textbf{69}}, 065212 (2004).


\bibitem{ven} G.M. Shore, G. Veneziano, Nucl. Phys. {\textbf{B381}}, 23 (1992)

\bibitem{feld} T. Feldmann, Int. J. Mod. Phys. A {\textbf{15}} 159 (2000); also available
as hep-ph/9907491.

\bibitem{emc} J. Ashman {\textit{et al.}}, Nucl. Phys. B
{\textbf{328}}, 1 (1989).

\bibitem{hirai} M. Hirai, S. Kumano, N. Saito, Phys. Rev. D  {\textbf{69}} 054021 (2004); 
also available as hep-ph/0312112.

\bibitem{elster_and_co} A. Sibirtsev, Ch. Elster, S. Krewald, and 
J. Speth, AIP Conf. Proc.{\textbf{717}}, 837 (2004); A. Sibirtsev, 
Ch. Elster, S. Krewald, and J. Speth, nucl-th/0303044.










\end{thebibliography}
\end{document}